\documentclass[12pt]{article}
\usepackage{epsfig}
\usepackage[usenames, dvipsnames]{color}
\usepackage{float} 
\usepackage{caption} 

\newcommand*\numu{\ensuremath{\nu_{\mu}~}}
\newcommand*\numubar{\ensuremath{\bar{\nu}_{\mu}~}}
\newcommand*\nue{\ensuremath{\nu_{e}~}}
\newcommand*\nuebar{\ensuremath{\bar{\nu}_{e}~}}
\usepackage{url}

\def\beq{\begin{equation}}
\def\eeq#1{\label{#1}\end{equation}}
\def\eeqn{\end{equation}}

\def\beqa{\begin{eqnarray}}
\def\eeqa#1{\label{#1}\end{eqnarray}}
\def\eeqan{\end{eqnarray}}

\let\bar=\overbar

\def\Dslash{\not{\hbox{\kern-4pt $D$}}}
\def\dslash{\not{\hbox{\kern-2pt $\del$}}}

\def\msb{{\bar{\ssstyle M \kern -1pt S}}}

\def\Title#1{\begin{center} {\Large {\bf #1} } \end{center}}

\begin{document}

\Title{Study of a New Target Design with an Additional Horn for NuMI Beam} 

\bigskip\bigskip


\begin{raggedright}  

{\it Jyoti Tripathi\index{Reggiano, D.}\\
Department of Physics, Panjab University\\
Chandigarh-160014, India\\
Fermilab, USA\\
On behalf of NOvA Collaboration}
\bigskip\bigskip
\end{raggedright}

{\center Talk presented at the APS Division of Particles and Fields Meeting (DPF 2017), July 31-August 4, 2017, Fermilab. C170731}

\section{Introduction}

NOvA is a long baseline neutrino experiment looking for \numu to \nue oscillations that utilizes NuMI (Neutrinos at the Main Injector) at Fermilab. The NuMI beamline has been constructed to deliver an intense \numu beam to various neutrino experiments for the Fermilab neutrino program. NuMI produces a tertiary neutrino beam by steering a 120 GeV proton beam onto a narrow 1.2 m long graphite target through a collimating baffle. Magnetic horns select the charged particles, mainly pions and kaons produced (positive for neutrinos and negative for antineutrinos) and focus them in the forward direction. Most of these particles subsequently decay into neutrinos in a 675 m long decay pipe. This paper describes the studies done for a new NOvA target design (also referred to as the new minimal target) and the placement of an addtional horn along the stripline to maximize the neutrino yield per proton in FHC (Forward Horn Current) as well as in RHC (Reverse Horn Current) configurations.

\section{Overview of the NOvA Standard Target and Horn Configuration}
The 1.2 m long NOvA target consists of 50 graphite segments (2.49 nuclear interaction length in total) as shown in Fig. 1. There are 48 graphite segments (fins) 24 mm long in the beam direction equally spaced with gaps of 0.5 mm between
them. The fins are 7.4 mm wide by 63 mm high and are cooled by water circulating in a baseplate at the
bottom of the fins. The beam is centered from each horizontal edge of the fin width and
3.7 mm below the top of the fin height. There are two graphite segments acting as Budal Monitors \cite{DUMMY:1} and electrically isolated from the rest of the target, one is vertical like the other fins to
measure the horizontal position of the beam and the other one is mounted horizontally to measure the vertical
position of the beam. The graphite is
ZXF-5Q from Poco Graphite and has a density of 1.78 g/cc. The NOvA fins are wide enough to accommodate a beam
spot of 1.3 mm RMS at 700 kW.  
\begin{figure}[H]                             
\centering
\includegraphics{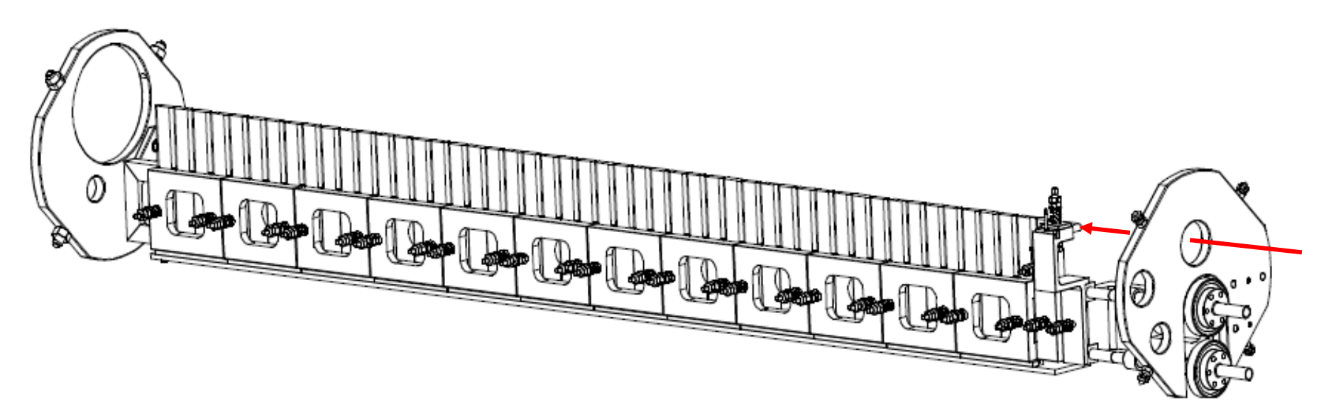}
\captionsetup{size=footnotesize}
\caption{Sketch of the NOvA Target consisting of 48 graphite segments \cite{DUMMY:4} and two Budal Monitors (right
end). The primary proton beam runs through the target from right to left (red arrow). The target is
surrounded by a cylinder (not shown) cooled by an additional water system. The ends of the cylinder have
two openings with thin beryllium windows. The upstream opening on the right end holds a beryllium
window 25.4 mm in diameter and 0.254 mm thick. The fin water cooling line enters and returns through
the bulkhead at the right end of the cylinder. At the downstream left end of the cylinder is another
beryllium window 120 mm in diameter and 1.2 mm thick which allows secondary particles to exit the
target cylinder with minimal additional interactions.}
\end{figure}
Downstream of the target, there is a focusing system composed of two magnetic horns with the Horn 1 placed at the
origin of the reference system (MCZERO) as shown in Fig. 2. There is a 5 cm stay clear area between the
downstream end of the target and the upstream end of Horn 1 since the
target is inserted blindly next to the Horn 1 inside the Target Hall. The downstream end of
the NOvA target is at -19.43 cm relative to MCZERO.
\begin{figure}[H]                          
\centering
\includegraphics[scale = 0.5]{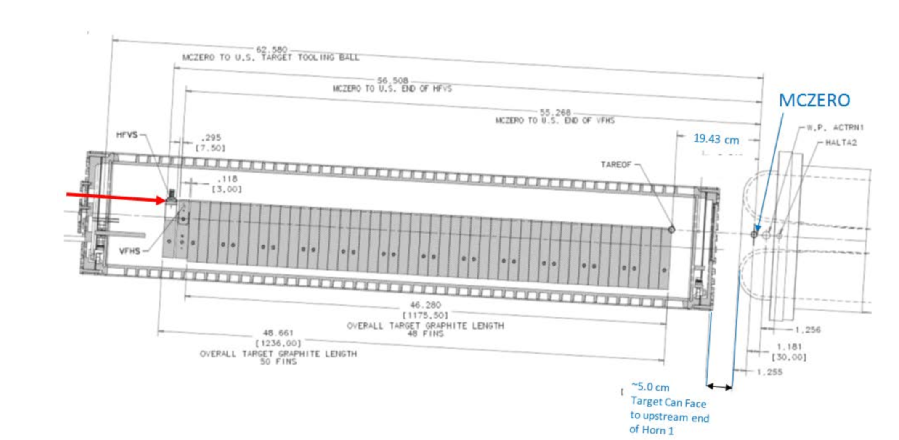}
\captionsetup{size=footnotesize}
\caption{Target distance from Horn 1 \cite{DUMMY:4}. The proton beam enters the target from the left in this diagram
(\textcolor{red} {red arrow}).}
\end{figure}
The Horn 1 has been placed at the origin of the reference system (MCZERO) whereas the Horn 2 for NOvA has been moved from the Low
Energy (LE) position (10.0 m from MCZERO) to the Medium Energy (ME) position (19.18 m from
MCZERO) (See Fig. 3).
\begin{figure}[H]                          
\centering
\captionsetup{size=footnotesize}
\includegraphics[scale = 0.5]{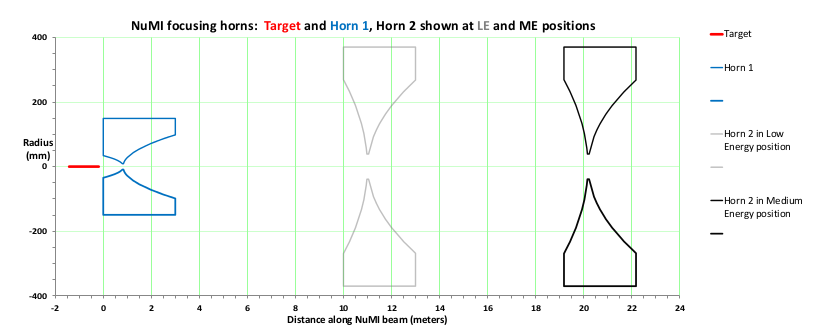}
\caption{NuMI target fins (red) as seen from above with the two focusing horns. Horn 1 (blue) starts at
MCZERO. Horn 2 is shown in two positions, the Medium Energy (ME, in black) and the Low Energy (LE,
in gray). The standard NOvA Horn 2 position is the ME position. Note that the vertical axis is in
millimeters, while the horizontal axis is in meters.}
\end{figure}


\section{Software Framework}

The beamline simulation for this study uses FLUGG simulation package \cite{DUMMY:2} that integrates Geant4 geometry description while using the physics models from FLUKA \cite{DUMMY:5} to simulate the interactions and decays of particles as they propagate.
The FLUGG simulation of the NuMI beamline begins with the primary proton beam, includes details of the beam line components like horns, decay pipe window, hadron absorber and muon monitors and models the hadronic cascade leading to a final neutrino flux.
For the beam simulation studies discussed in the upcoming sections, the versions used are as follows: 
\begin{itemize}
 \item FLUGG 2009\_3 
 \item FLUKA.2011.2c.4  
  \item  Beam spot size was kept at 1.1 mm. 
  \item  Horn current was set to 200 kA. 
  \item  All the plots are scaled to 6e20 protons on target (POTs).
\end {itemize}

\section{The New NOvA Minimal Target Design and Horn Configuration}
The conceptual design of the new minimal target consists of two components: a 24 fin upstream (exactly like the standard NOvA target) and a 24 fin downstream part (with the new design because
it gets inserted inside of Horn 1). We chose to leave the upstream 24 fin design unchanged because the standard NOvA target has
been a robust design during NuMI running to date. A good design principle is to
keep the cooling function as far as possible from the graphite fin target portion hit by the NuMI beam.
As the new minimal target design would see higher exposure in the coming years, this assumption could
change as experience is acquired.
\begin{figure}[H]                        
\centering
\includegraphics[scale = 0.5, angle = -90]{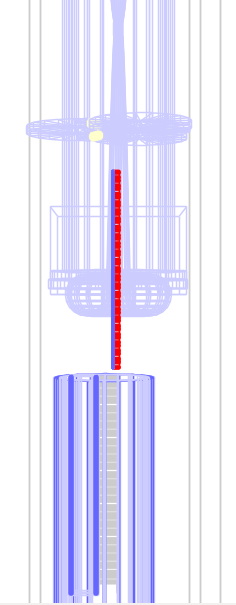}
\captionsetup{size=footnotesize}
\caption{The new NOvA minimal target with fins pushed inside Horn 1 (red).}
\end{figure}
The new minimal target design does have one additional advantage. The downstream part of the target that goes inside
Horn 1 is only 60 cm long (the part inside of Horn 1 is only 46.55 cm long). The MINOS target design
was 94 cm long and had very little strength in the horizontal direction. If the full 120 cm long
new minimal target was all one design like the MINOS target, it would be fragile in the horizontal
direction.
The cross-sectional view of the new minimal target is shown in Fig 5. The intent is to
maximize the bare fin distance between the beam spot and the cooling structure. The height of these 24 fins has been reduced to 17 mm from 24 mm to fit inside the Horn 1. The radius of the surrounding circular shell is 21.5 mm. At the maximum insertion point (+40 cm relative to MCZERO) the inner conductor inner surface of the Horn 1 is at 25.4 mm, so there is about 4 mm of
radial clearance between this target and the Horn 1.
\begin{figure}[H]                        
\centering
\includegraphics[scale = 0.5]{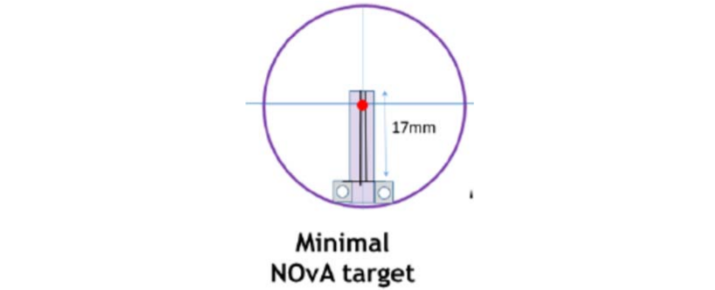}
\captionsetup{size=footnotesize}
\caption{Cross-Sectional view of the new NOvA minimal target}
\end{figure}

\section{The Event yield with the New NOvA Minimal Target}
Table 1 and Table 2 show the total \numu and \numubar event yield in ME configuration for the standard and the new NOvA minimal target respectively. For the \numu flux, the new minimal target gains 11.5\% relative to the standard target at the Near Detector and 11.3\% at the Far Detector respectively.

\begin{table}[H]
\centering
\captionsetup{font=footnotesize}
\caption{\numu event yield and the \nue contamination for standard and new minimal target for FHC in 1-3 GeV energy range. For Far Detector, the events have been multiplied by a factor of $10^6$.}
\begin{tabular}{l|cccc}  
Target Design    &  ND \numu    &  FD \numu     &   ND \nue   & FD \nue  \\ \hline
Standard Target  &   79.41     &     91.51      &     0.54    &  0.63  \\
New Minimal Target  &   88.60      &     101.80      &     0.57    &  0.65   \\ \hline
Gain             &  11.57\%       &     11.24\%      &     5.55\%   &  3.17\%  \\ \hline
\end{tabular}
\end{table}

\begin{table}[H] 
\centering
\captionsetup{font=footnotesize}
\caption{\numubar event yield and the \nuebar contamination for standard and new minimal target for RHC in 1-3 GeV energy range. For Far Detector, the events have been multiplied by a factor of $10^6$.}
\begin{tabular}{l|cccc}  
Target Design    &  ND \numubar   &  FD \numubar      &   ND \nuebar   & FD \nuebar \\ \hline
Standard Target  &   30.17     &     34.75      &     0.20    &  0.22  \\
New Minimal Target  &   33.93     &     38.94      &     0.22    &  0.26   \\ \hline
Gain             &  12.46\%       &     12.05\%      &     10\%   &  18.18\%  \\ \hline
\end{tabular}

\end{table}
\subsection{Optimization of the Horn 2 Position}
The target to horn distance is flexible and the separation between two horns can also be changed. 
The inelastic collisons which produce mesons gives a transverse momentum to the particles produced, peaking at 0.3 GeV/c, with only slight dependence on the meson longitudinal momentum. The typical production angle of the mesons is inversely proportional to meson momentum. Hadrons produced in the target along the beam axis pass through the horns unaffected. Hadrons well focused by Horn 1 pass unaffected by the Horn 2. Horn 2 essentially increases the efficiency of the focusing system by focusing a significant fraction of the hadrons not well focused by Horn 1.
The position of Horn 2 with respect to Horn 1 along the stripline optimizes the focusing of the pions already focused by the Horn 1. We observed that for a few positions of Horn 2, the event yield is slightly better than that obtained at 19 m. We repeated the same study for the new minimal target where we scanned different positions of Horn 2 in steps of 1 m, beginning from 6 m and going upto 19 m w.r.t MCZERO. 
\begin{figure}[H]
\centering
\captionsetup{font = footnotesize}
\includegraphics[scale = 0.39]{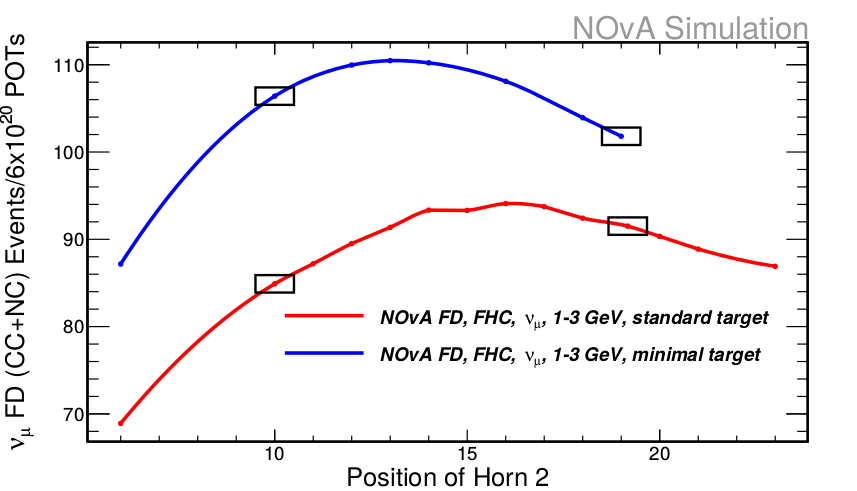}
\caption{\numu event numbers for different positon of Horn 2 with respect to Horn 1 in FHC for standard (red) and new minimal (blue) NOvA Target in 1-3 GeV energy range.}
\end{figure}
\begin{figure}[H]                        
\centering
\captionsetup{font = footnotesize}
\includegraphics[scale = 0.33]{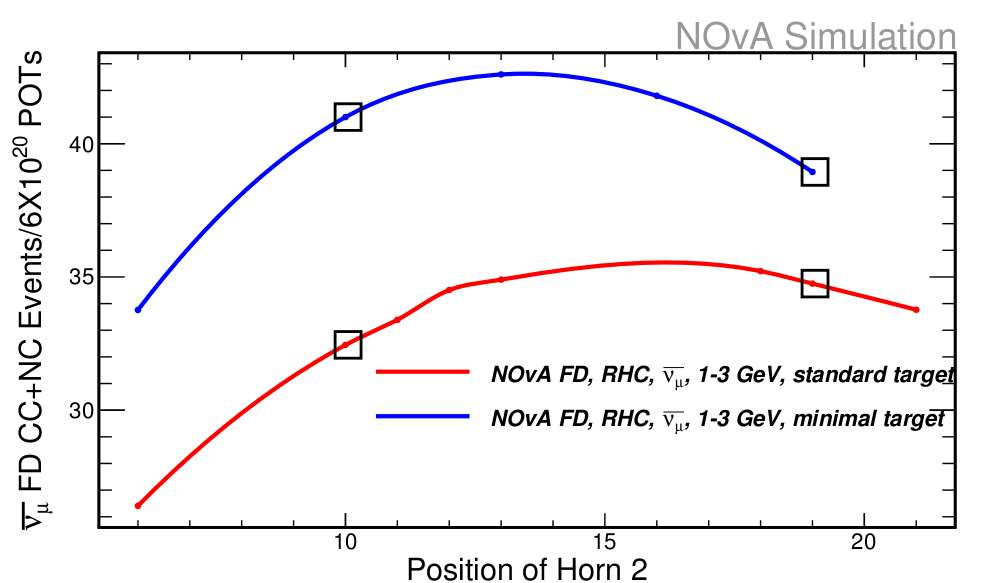}
\caption{\numubar event numbers for different position of Horn 2 w.r.t to Horn 1 in RHC  for standard (red) and new minimal (blue) NOvA target in the 1-3 GeV energy range.}
\end{figure}
The current target with the LE configuration is less productive than the ME one, for both FHC and RHC, however it is quite the opposite with the new minimal target design where the LE produces more yield than the ME configuration.
Another interesting observation is that the new minimal target gives maximum number of \numu events with the Horn 2 placed at 13 m. The \nue contamination also goes up but a very small fraction.
\begin{table}[H]
\centering
\captionsetup{font = footnotesize}
\caption{\numu event yield and the \nue contamination for standard target (ME configuration) and new minimal target (Horn 2 placed at 13 m) for FHC in 1-3 GeV energy range. For Far Detector, the events have been multiplied by a factor of 10$^6$. The last row shows the percentage gain in the neutrino flux relative to that obtained with the standard target.}
\begin{tabular}{l|cccc}  
Target Design    &  ND \numu    &  FD \numu     &   ND \nue   & FD \nue \\ \hline
Standard Target  &   79.41     &     91.51      &     0.54    &  0.63  \\
New Minimal Target  &  97.41      &     110.46      &     0.70    &  0.82   \\ \hline
Gain             &  22.67\%       &     20.71\%      &     29.62\%   &  30.16\%  \\ \hline
\end{tabular}

\end{table}
\begin{table}[H]
\centering
\captionsetup{size=footnotesize}
\caption{\numubar event yield and the \nuebar contamination for standard target(ME configuration) and new minimal target (Horn 2 placed at 13 m) for RHC in 1-3 GeV energy range. For Far Detector, the events have been multiplied by a factor of 10$^6$. The last row shows the percentage gain in the anti-neutrino flux relative to that obtained with the standard target.}
\begin{tabular}{l|cccc}  
Target Design    &  ND \numubar    &  FD \numubar     &   ND \nuebar   & FD \nue \\ \hline
Standard Target  &   30.17     &     34.75      &     0.20    &  0.22  \\
New Minimal Target  &  37.62      &     42.68      &     0.26    &  0.30   \\ \hline
Gain             &  24.69\%       &     22.82\%      &     30\%   &  36.36\%  \\ \hline
\end{tabular}

\end{table}

\begin{figure}[H]  
\centering
\includegraphics[scale = 0.3]{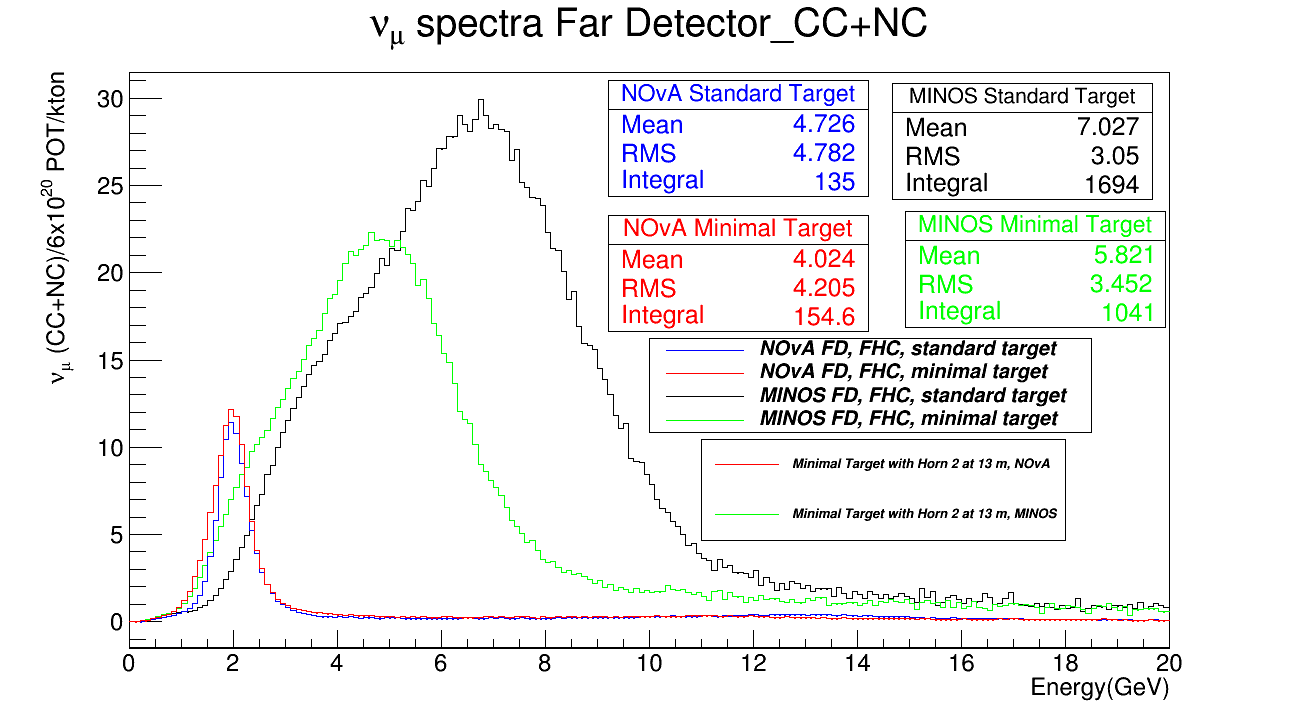}
\captionsetup{size = footnotesize}
\caption{\numu energy spectra for new minimal target with Horn 2 at 13 m for NOvA (in red) and MINOS (in green) at the Far Detector for FHC. The spectra corresponding to the standard target in ME configuration are in blue (NOvA) and black (MINOS).}
\end{figure}


\begin{figure}[H] 
\centering
\includegraphics[scale = 0.3]{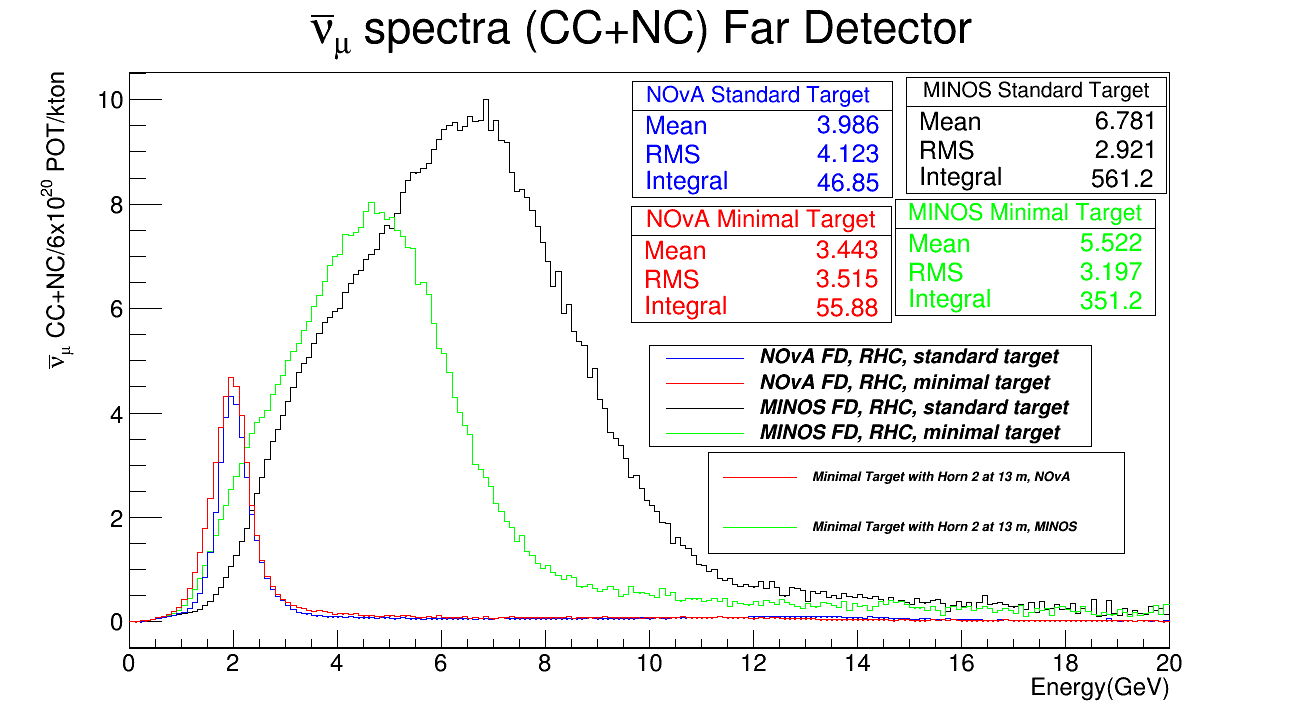}
\captionsetup{size = footnotesize}
\caption{\numubar energy spectra for new minimal target with Horn 2 at 13 m for NOvA (in red) and MINOS (in green) at the Far Detector for RHC. The spectra corresponding to the standard target in ME configuration are in blue (NOvA) and black (MINOS).}
\end{figure}

\subsection{Introduction of an Additional Horn}
Our simulation studies have shown that the Horn 1 by itself focuses 60-65\% of the neutrino yield.  Introduction of an additional horn helps focusing pions exiting at large angles from the Horn 1. We simulated the three horn configuration with the introduction of an additional horn, identical to the Horn 2 since an identical Horn 2 might help avoid the long lead time for engineering and prototyping of a completely new horn design.  
\begin{figure}[H] 
\centering  
\includegraphics[scale = 0.3]{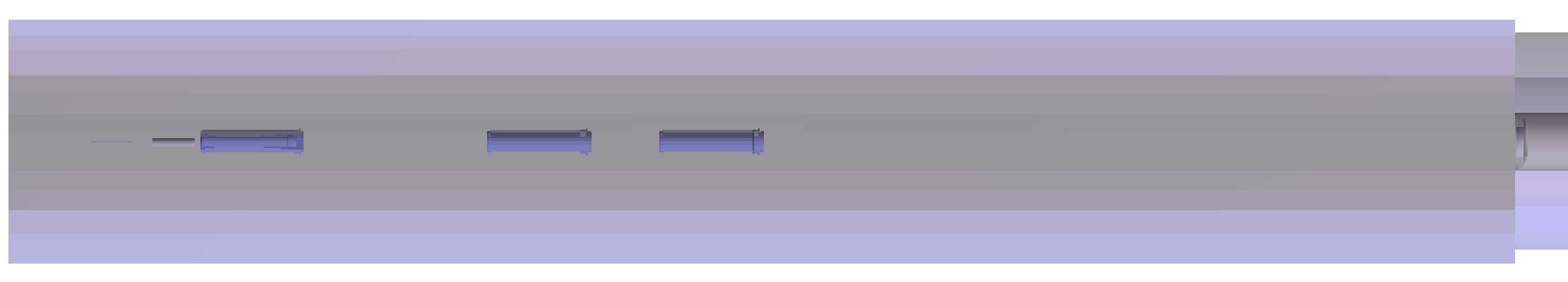}
\captionsetup{size=footnotesize}
\caption{Three horn configuration with two identical Horn 2s along the striplines}
\end{figure}
The placement of a Horn 2 both at low energy and medium energy positions might be helpful in refocusing low energy pions that have been overfocused by Horn 1. We investigated how the neutrino flux changes when we move the position of the two Horn 2s relative to each other and relative to Horn 1 for the new minimal target design.
\begin{figure}[H]  
\centering
\includegraphics[scale = 0.6]{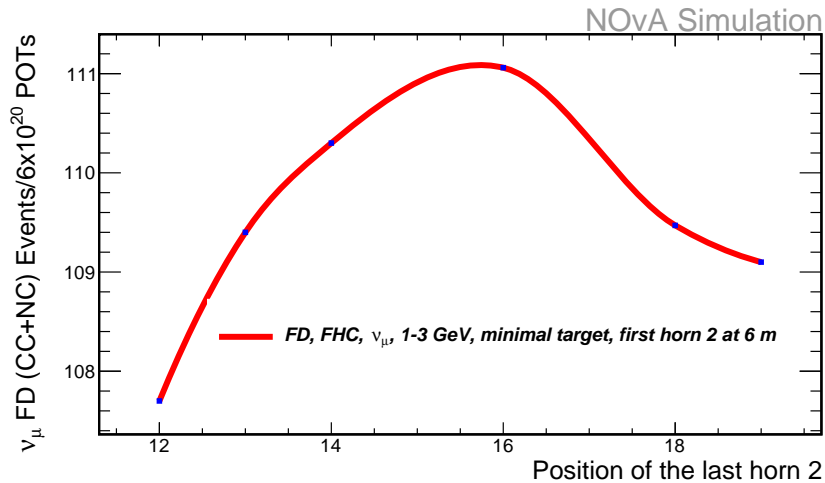}
\captionsetup{size=footnotesize}
\caption{\numu event yield at the Far Detector for scanning the position of the second horn fixing the first horn 2 at 6 m for FHC configuration for the new minimal target.}
\end{figure}
\begin{table}[H]
\centering
\captionsetup{size=footnotesize}
\caption{\numu event yield and the \nue contamination for standard target (ME configuration) and new minimal target (with two Horn 2s placed at 6 m and 16 m) for FHC in 1-3 GeV energy range. For Far Detector, the events have been multiplied by a factor of 10$^6$. The last row shows the percentage gain in the neutrino flux relative to that obtained with the standard target.}
\begin{tabular}{l|cccc}  
Target Design    &  ND \numu    &  FD \numu     &   ND \nue   & FD \nue \\ \hline
Standard Target  &   79.41     &     91.51      &     0.54    &  0.63  \\
New Minimal Target  &  96.64      &     111.06      &    0.69    &  0.79   \\ \hline
Gain             &  21.69\%    &     21.36\%     &     27.77\%   &  25.39\%  \\ \hline
\end{tabular}

\end{table}The \numu event yields produced for different positions of the two horn 2s were recorded (see Fig. 11). We observed a maximum number of \numu events when the first Horn 2 is placed at 6 m and the last one at 16 m.

\section{Summary}
Our simulation studies show that moving the target fins inside Horn 1 in the new minimal target design is more productive than the standard target and we gain more neutrino yield without any shift in the peak position of the neutrino spectra. However moving the target inside the Horn 1 may increase the stress on the target material. This is being studied by the engineers from the Accelerator Division. An additional Horn identical to Horn 2 shows promising results. A detailed optimization of the position of the additional Horn is underway with the new target design. The new target design would have a significant impact on the time required to reach the desired significance for the oscillation parameters that NOvA is studying.
\bigskip

\section{Acknowledgement}
NOvA is supported by the US Department of Energy; the US National Science Foundation;the Department of Science and Technology, India; the European Research Council; the MSMT CR, Czech Republic; the RAS, RMES, and RFBR, Russia; CNPq and FAPEG, Brazil; and the State and University of Minnesota. We are grateful for the contributions of the staffs of the University of Minnesota module assembly facility
and NOvA FD Laboratory, Argonne National Laboratory, and Fermilab. Fermilab is operated by Fermi Research Alliance, LLC under Contract No. DeAC02-07CH11359 with the US DOE.
\bibliographystyle{unsrt}  
\nocite{*}
\bibliography{reference}
\bibliographystyle{ieeetr} 
\end{document}